\documentclass[sigconf,nonacm,screen]{acmart}

\usepackage{siunitx}
\usepackage[most]{tcolorbox}
\usepackage{listings, color}
\usepackage{hyperref}
\usepackage[hyphenbreaks]{breakurl}
\usetikzlibrary{positioning}

\usepackage[normalem]{ulem}
\usepackage{csquotes}
\usepackage{booktabs}
\usepackage{paralist}
\usepackage{graphics}
\usepackage{makecell}

\usepackage{multirow}
\usepackage{colortbl}
\usepackage{subcaption}
\usepackage{amsthm}

  {\small \em \tt ccc\begin{table} aaa bbb}%
  {\end{table} ddd}

\newcommand{\json}{JSON}
\newcommand{\kw}[1]{\text{\bf #1}}
\renewcommand{\kw}[1]{\ensuremath{\mathtt{#1}}}

\newcommand{\akey}[1]{\ensuremath{\mathsf{#1}}}

\newcommand{\xnot}{\kw{not}}

\newcommand{\xone}{\kw{oneOf}}

\newcommand{\xany}{\kw{anyOf}}

\newcommand{\xall}{\kw{allOf}}

\newcommand{\xreq}{\kw{required}}

\newcommand{\xtype}{\kw{type}}

\newcommand{\xprops}{\kw{properties}}

\newcommand{\xpattProps}{\kw{patternProperties}}

\newcommand{\xaddProps}{\kw{additionalProperties}}

\newcommand{\xpatt}{\kw{pattern}}

\newcommand{\xit}{\kw{items}}

\newcommand{\xenum}{\kw{enum}}

\newcommand{\xconst}{\kw{const}}

\newcommand{\xdref}{\kw{\$ref}}

\newcommand{\xdcomm}{\kw{\$comment}}

\newcommand{\xdescr}{\kw{description}}

\newcommand{\xtitle}{\kw{title}}

\newcommand{\gcomment}[1]{}
\newcommand{\hide}[1]{}
\newcommand{\code}[1]{}

\newcommand{\Num}{\akey{Num}}
\newcommand{\Str}{\akey{Str}}

\newcommand{\GG}[1]{}

\newcommand{\ST}[1]{}

\newcommand{\AB}[1]{}

\newlength{\NL}
\newcommand{\nule}{\nullv}
\newcommand{\n}{n}
\newcommand{\nullv}{\ensuremath{\text{null}}}
\newcommand{\str}{s}

\newcommand{\J}{{J}}

\newcommand{\truev}{\ensuremath{\text{true}}}
\newcommand{\falsev}{\ensuremath{\text{false}}}

\newcommand{\jsonsch}{JSON Schema} 

\definecolor{burntorange}{rgb}{0.8, 0.33, 0.0}
\definecolor{mygray}{rgb}{0.643,0.643,0.643}
\definecolor{cerulean}{rgb}{0.0, 0.48, 0.65}
\newtcolorbox{schemabox}[2][]{%
  sidebyside align=top,
  enhanced,
  boxsep=0pt,
  arc=0pt,
  top=-3pt, bottom=-3pt,
  left=2pt, right=0pt,
  colback=mint!10,
  colframe=mint,
  boxrule=1.2pt,
} 

\theoremstyle{remark}
\newtheorem{theorem}{Theorem}
\newtheorem{example}[theorem]{Example}
\newtheorem{definition}[theorem]{Definition}

\definecolor{caribbeangreen}{rgb}{0.0, 0.8, 0.6}
\definecolor{dollarbill}{rgb}{0.52, 0.73, 0.4}
\definecolor{mint}{rgb}{0.24, 0.71, 0.54}

\newcommand{\webref}[2]{\href{#1}{#2}}
\usetikzlibrary{calc, arrows.meta, shapes, matrix, scopes, backgrounds}

\begin{document}
	\copyrightyear{2022}
	\title{Extracting JSON Schemas with Tagged Unions}
	
	\author{Stefan Klessinger}
	\affiliation{%
		\institution{University of Passau}
		\city{Passau}
		\state{Germany}
	}
	\email{stefan.klessinger@uni-passau.de}
	
	\author{Meike Klettke}
	\orcid{0000-0003-0551-8389}
	\affiliation{%
		\institution{University of Regensburg}
		\city{Regensburg}
		\country{Germany}
	}
	\email{meike.klettke@ur.de}
	
	\author{Uta Störl}
	\orcid{0000-0003-2771-142X}
	\affiliation{%
		\institution{University of Hagen}
		\city{Hagen}
		\country{Germany}
	}
	\email{uta.stoerl@fernuni-hagen.de}
	
	\author{Stefanie Scherzinger}
	\affiliation{%
		\institution{University of Passau}
		\city{Passau}
		\country{Germany}
	}
	\email{stefanie.scherzinger@uni-passau.de}
	
	\begin{abstract}
		With data lakes and schema-free NoSQL document stores, extracting a descriptive schema from JSON data collections 
		is an acute challenge.
		In this paper, we target the discovery of tagged unions, a JSON Schema design pattern where the value of one property of an object (the tag) conditionally implies subschemas for sibling properties.
		We formalize these implications as conditional functional dependencies and capture them using the JSON Schema operators if-then-else. 
		We further motivate our heuristics to avoid overfitting.
		Experiments with our prototype implementation are promising, and show that this form of tagged unions can successfully be detected in real-world GeoJSON and TopoJSON datasets. In discussing future work, we outline how our approach can be extended further.
	\end{abstract}
	\maketitle
	
	\section{Introduction}
	
	JSON is a popular data exchange format. Extracting  a schema from collections of JSON documents is a real-world challenge which is actively being researched~\cite{DBLP:conf/btw/KlettkeSS15,DBLP:journals/vldb/BaaziziCGS19,DBLP:conf/sigmod/SpothKLHL21,DBLP:conf/er/ContosS20,enase22,klettke2017uncovering,DBLP:conf/edbt/BaaziziBCGS20,DBLP:conf/vldb/Namba21}.
	
	Ideally, the extracted schema describes the data \emph{tightly}, yet without overfitting. 
	In this article, we  target the detection of a specific schema design pattern in the JSON Schema language, the pattern of \emph{tagged unions}, also known as discriminated unions, labeled unions, or variant types. 
	This is a recommended design pattern~\cite{jsifthenelse}, and has been found to be quite common in real-world schemas~\cite{DBLP:conf/er/BaaziziCGSS21}.
	
	\makeatletter
	\newcommand\querysize{\@setfontsize\querysize\@vipt\@viipt}
	\makeatother
	
	\lstdefinestyle{query}{
		numbers=left,
		stepnumber=1,
		numbersep=3pt,
		xleftmargin=7pt,
		tabsize=4,
		showspaces=false,
		showstringspaces=false,
		basicstyle=\linespread{1}\fontfamily{lmtt}\selectfont\querysize,
		keywordstyle=\color{blue},
		stringstyle=\color{purple},
		upquote=true,
		breaklines=true,
		commentstyle=\color{CadetBlue},
		moredelim=[is][\textit]{[*}{*]}
	}
	
	\definecolor{mygray}{rgb}{0.643,0.643,0.643}
	\newtcolorbox{queryboxlink}[3][]{%
		sidebyside align=top,
		enhanced,
		boxsep=0pt,
		arc=0pt,
		top=-3pt, bottom=-3pt,
		left=2pt, right=0pt,
		colback=mint!10,
		colframe=mint,
		boxrule=1.2pt,
	} 
	
	\begin{example} \label{ex:geojson}
		Figure~\ref{fig:geojson_data} shows GeoJSON data. The array starting in line~2
		holds four objects, of type {\tt Point} and {\tt LineString}.
		Point coordinates are encoded as an array of numbers, while lines are encoded as an array of points, and hence, an array of number arrays.
		
		\begin{figure}[th!]
			\centering
			\begin{queryboxlink}{}{}
				\begin{lstlisting}[style=query]   
					{ "type": "GeometryCollection",
						"geometries": [
						{ "type": "Point",
							"coordinates": [30,10] },
						{ "type": "Point",
							"coordinates": [40,15] },
						{ "type": "LineString",
							"coordinates": [[55,5], [10,30], [10,10]] },
						{ "type": "LineString",
							"coordinates": [[30,10], [10,30], [40,40]] }
						] }
				\end{lstlisting}
			\end{queryboxlink}  
			
			\caption{GeoJSON data. The value of property {\tt type} (here {\tt Point} and {\tt LineString})
				determines the subschema of property {\tt coordinates} as either  encoding a geometric point (an array of numbers), or a line (an array of points).}
			\label{fig:geojson_data}
		\end{figure}
		The GeoJSON  specification (\webref{https://datatracker.ietf.org/doc/html/rfc7946}{IETF RFC 7946}, clickable link embedded in the PDF) describes six types of geometries, including polygons and  multi-polygons.
		Consistently, the property {\tt type}  serves as a \emph{tag} to distinguish the subschema of sibling property \verb!coordinates!, thereby instantiating a tagged union. 
		
		GeoJSON comes without an official JSON Schema specification~\cite{DBLP:journals/geoinformatica/FrozzaM20}. In Figure~\ref{fig:geojson_schema}, we therefore show a hand-crafted excerpt using the \verb!if-then-else! construct to enforce the tagged union, with the following semantics: If the object in question has a property labeled {\tt type} with the value {\tt Point}, then the value of property {\tt coordinates} must be an array of numbers.
		Else, if the property {\tt type} has the value {\tt LineString}, the value of {\tt coordinates} must be an array of points, and hence, an array of number arrays.
	\end{example}

	\begin{figure}[thb]
		\centering
		\begin{queryboxlink}{}{}
			\begin{lstlisting}[style=query]   
				"if": {
					"properties": {
						"type": { "const": "Point" } },
					"required": [ "type" ] },
				"then": {
					"properties": {
						"coordinates": {
							"type": "array",
							"items": { "type": "number" } } } },
				"else": {
					"if": {
						"properties": {
							"type": { "const": "LineString" } },
						"required": [ "type" ] },
					"then": {
						"properties": {
							"coordinates": {
								"type": "array",
								"items": {
									"type": "array",
									"items": { "type": "number" } } } } } }
			\end{lstlisting}
		\end{queryboxlink}  
		
		\caption{JSON Schema snippet declaring a tagged union for GeoJSON geometry objects {\tt Point} and {\tt LineString}.
		}
		\label{fig:geojson_schema}
	\end{figure}
	
	\begin{example}  \label{ex:other_datasets}
		Various schemas listed on \webref{https://github.com/SchemaStore/schemastore}{SchemaStore.org}, 
		a com\-munity-curated collection of JSON Schema declarations,
		encode tagged unions using
		\verb!if-then-else!.
		Among them are the \webref{https://github.com/SchemaStore/schemastore/blob/6049f681741399cf73aea088680f58375c12592a/src/schemas/json/minecraft-predicate.json}{Minecraft} schemas 
		which use tagged unions to encode so-called data packs for configuring Minecraft worlds.
		Other examples are \webref{https://github.com/SchemaStore/schemastore/blob/6049f681741399cf73aea088680f58375c12592a/src/schemas/json/github-issue-forms.json}{Github Issue Forms}, or the 
		\webref{https://github.com/SchemaStore/schemastore/blob/6049f681741399cf73aea088680f58375c12592a/src/schemas/json/cloudify.json}{cloudify} schema (clickable links in the PDF). 
	\end{example}
	
	In schema extraction, the detection of tagged unions has so far received limited attention:
	Several approaches are able to detect \emph{union types}, i.e., properties whose type is a union of types,
	but they do not detect the dependency w.r.t.\ a specific property value (the \enquote{tag} in the tagged union). 
	Figure~\ref{fig:union_types_a} shows an example: This schema for GeoJSON data does not restrict the values for property {\tt type} (it allows all strings) and allows alternative subschemas for property {\tt coordinates}. As it is overly general, the schema allows to encode lines as mere arrays of numbers (rather than arrays of number-arrays), in violation of the GeoJSON specification.
	
	Existing approaches to JSON Schema extraction fail to detect tagged unions.
	Notably, in describing their approach to schema extraction based on typing JSON values, Baazizi et al.~\cite{DBLP:journals/vldb/BaaziziCGS19} outline how their approach can be extended to include tagged unions.
	However, they do not discuss strategies against overfitting to the input data during the discovery of tagged unions. 
	
	Our proposal follows a different approach, and relies on a relational encoding of JSON objects from which we then derive conditional dependencies. A central part of our contribution are our heuristics, which filter out dependencies that have insufficient support in the input data, so that they are not reflected in the derived schema. Specifically, ours is the first proposal towards the discovery of tagged unions that --- to our knowledge --- includes an  experimental evaluation over real-world data.

	\paragraph{Contributions.} This paper makes the following contributions:
	
	\begin{compactitem}
		\item We target the detection of tagged unions in JSON schema extraction, specifically, tagged unions that are based on dependencies between a property value and the implied subschema of its sibling properties.
		
		\item Our approach relies on the discovery of unary constant conditional functional dependencies in a relational encoding of the JSON objects.
		Traditionally, conditional functional dependencies are employed in the context of data cleaning, and we apply them to a new domain.
		
		\item Our approach is composable with existing algorithms for JSON schema extraction, as we impose the tagged unions on top of the schemas derived by state-of-the-art third-party tools.

		\item Our experiments on real-world GeoJSON datasets (and its sibling format TopoJSON) show that meaningful tagged unions can indeed be identified.
		We illustrate the impact of a configurable threshold on the number of tagged unions detected, and consequently, the size of the extracted schema.
		
	\end{compactitem}
	
	We further outline promising directions of future work.

	\paragraph{Artifact availability and reproducibility.}
	We have made our research artifacts (code, data, extracted schemas), as well as a fully automated reproduction
	package, long-term available online~\cite{zenodo:repro}.

	\paragraph{Structure of this paper.}
	In Section~\ref{sec:preliminiaries}, we introduce the preliminaries. Section~\ref{sec:related_work} discusses related work. In Section~\ref{sec:approach}, we present our approach, specifically the architecture and heuristics employed, and an outlook on future extensions. In Section~\ref{sec:experiments}, our experiments are presented and discussed. In Section~\ref{sec:conclusion}, we draw our conclusions and discuss opportunities for future work. 
	
	\section{Preliminaries}
	\label{sec:preliminiaries}
	
	\subsection{JSON Data Model and JSON Schema}
	
	\paragraph{JSON data model.}
	\mbox{}\\
	The grammar below (adopted from~\cite{DBLP:journals/corr/abs-2202-12849}) captures the syntax of JSON values,
	namely basic values, objects, or arrays.
	Basic values~$B$ include  the \nule\ value, Booleans, numbers~$\n$, and strings~$\str$.
	Objects~$O$ represent sets of members, each member being a name-value pair, 
	arrays~$A$ represent sequences of values.
	{\small
		\[
		\begin{array}{@{} l @{\hskip 3pt} llr @{\hskip 0pt} l}
			\J ::= 
			& B \mid O \mid A                       
			& 
			\\
			B ::=	 
			& \nule \mid \truev \mid \falsev \mid \n \mid \str  \quad \quad
			& \n\in\Num, \str\in\Str  
			\\
			O ::= 
			&  \{l_1:\J_1,\ldots,l_n:\J_n \}\,     
			& n\geq 0, \ \  i\neq j \Rightarrow l_i\neq l_j
			\\
			A ::= 
			&  [\J_1, \ldots, \J_n ]\, 
			& n\geq 0  
			\\
		\end{array}
		\]
	}
	
	\paragraph{JSON Schema.} \label{sec:jsonschema}
	{\jsonsch} is a language for defining the structure of {\json} documents. 
	The syntax and semantics of {\jsonsch} have been formalized in~\cite{DBLP:conf/www/PezoaRSUV16}, 
	and we informally present some of the main keywords:
	
	Assertions include \xreq, \xenum, \xconst, \xpatt\ and \xtype, and indicate a test that is performed on the
	corresponding instance.
	
	Applicators include the Boolean operators \xnot, \xany, \xall, \xone, as well as,  \verb!if-then-else!. They further include the object operators \xpattProps, \xaddProps, and \xprops, and the array operator \xit,
	and the reference operator \xdref. They apply a different 
	operator to the same instance or to a component of the current instance.
	
	Annotations (\xtitle, \xdescr, and \xdcomm) do not affect validation
	but indicate
	an annotation associated to the instance.
	\begin{figure*}
\begin{subfigure}[t]{0.49\textwidth}
\begin{schemabox}{Klettke et al}
\lstset{escapeinside={(*}{*)}}
\begin{lstlisting}[style=query]  
"anyOf": [
  { "type": "object", 
    "properties": {
      (*\textcolor{burntorange}{"type": \{ "type": "string" \},}*)
      "coordinates": { 
        "type": "array",
        "items": { "type": "number" } } } 
  },
  { "type": "object", 
    "properties": {
      (*\textcolor{burntorange}{"type": \{ "type": "string" \},}*)
      "coordinates": { 
        "type": "array",
        "items": { "type": "array",
          "items": { "type": "number" }
   } } } } ]
\end{lstlisting}
\end{schemabox}
\caption{Union type.}
\label{fig:union_types_a}
\end{subfigure}
\begin{subfigure}[t]{0.49\textwidth}
\begin{schemabox}{Frozza et al}
\lstset{escapeinside={(*}{*)}}
\begin{lstlisting}[style=query]   
"anyOf": [
  { "type": "object", 
    "properties": {
      (*\textcolor{burntorange}{"type": \{ "const": "Point" \},}*)
      "coordinates": { 
        "type": "array",
        "items": { "type": "number" } } } 
  },
  { "type": "object", 
    "properties": {
      (*\textcolor{burntorange}{"type": \{ "const": "LineString" \},}*)
      "coordinates": { 
        "type": "array",
        "items": { "type": "array",
          "items": { "type": "number" }
   } } } } ]
\end{lstlisting}
\end{schemabox}
\caption{Tagged union.}
\label{fig:union_types_b}
\end{subfigure}
\caption{Snippets of different encodings of GeoJSON geometries in JSON Schema. Left: Union type encoding, as extracted by traditional schema extraction tools (with syntactic variations, depending on the tool).  Right: Tagged union encoding using {\tt anyOf}, as an  alternative to the {\tt if-then-else} construct.
While near-identical in syntax, differing only in lines~4 and~11, the schema semantics differ, as the union type allows incorrect encodings of points or line strings in GeoJSON.}
\label{fig:third-party}
\end{figure*}
	\paragraph{Union types}
	A property with several possible types (or subschemas) can be described as a \emph{union type}~\cite{union_types}, i.e.,  the union of all types the property assumes. In JSON Schema, this can be encoded by a disjunction, using the union operators {\tt anyOf} or {\tt oneOf} (where the former is inclusive, and the latter exclusive). Such an encoding is exemplified in Figure~\ref{fig:union_types_a} for GeoJSON. 
	Union types are recognized by most of the existing tools for JSON Schema extraction (as we will discuss in greater detail in our discussion of related work). While the schema distinguishes two variants for encoding {\tt coordinates} (an array of numbers, or an array thereof), it does not capture any dependencies between the value of property {\tt type} and the subschema for the sibling property {\tt coordinates}. Also, the domain of property {\tt type} is not restricted to specific string values.

	\paragraph{Tagged unions}
	The {\tt if-then-else} operator allows for declaring tagged unions, and was introduced (rather recently) with JSON Schema Draft~7. 
	
	In Example~\ref{ex:geojson}, we informally introduced the semantics. As for terminology, we distinguish one property as the \emph{tag} (in our example, the property labeled {\tt type}), and identify one or more properties with \emph{implied subschemas} (in our example, for {\tt coordinates}). 
	
	The tag-property may go by any name. While the schemas for GitHub Issue Forms and cloudify
	from Example~\ref{ex:other_datasets}
	incidentally also rely on a tag labeled \enquote{type}, in the example of Minecraft (see data snippets in Figure~\ref{fig:minecraft_values}), the tag-property is labeled \enquote{condition}.

	An alternative encoding for tagged unions is to use a union operator, as shown in Figure~\ref{fig:union_types_b}.
	Two different combinations of values for property {\tt type} and subschemas for property {\tt coordinates} are given in accordance with the GeoJSON definition. Again, an object may either be a \enquote{Point} with coordinates encoded as an array of numbers or a \enquote{LineString} with an array of points. 
	
	Note that while Figures~\ref{fig:union_types_a} and~\ref{fig:union_types_b} differ only marginally in their syntax, the difference in semantics is striking: The tagged union captures the dependency between the value of property {\tt type} and the subschema of property {\tt coordinates}.
	While union type encodings can be derived by several state-of-the-art schema extraction tools, schemas with tagged unions are -- so far -- manually crafted, since existing approaches to schema extraction are not capable of discovering value-based dependencies.
	
	Our approach can produce either
	the {\tt if-then-else} encoding or the encoding exemplified in Figure~\ref{fig:union_types_b}. We chose to implement the former, because it is more lenient w.r.t.\ unexpected tag values (in this case no restrictions are specified). Note that this is not a limitation of our approach, and a merely technical limitation.

	\subsection{Dependencies}
	
	For the relational model, the concept of functional dependencies is well ex\-plored, and various generalizations are known, such as conditional functional dependencies~\cite{DBLP:conf/icde/BohannonFGJK07} which only apply to a subset of the tuples.
	In the following, we extend these notions to JSON data, assuming a relational encoding of all objects that are reachable by the same path from the document root. This idea of a relational encoding of semi-structured data for the definition or detection of dependencies is a common approach, e.g., for XML~\cite{10.1145/974750.974757}, JSON~\cite{DBLP:journals/corr/Mior2021}, or RDF data~\cite{DBLP:conf/sigmod/0001JPKQN16}.

	\paragraph{Relational encoding}
	Given a JSON value, we consider all labeled paths from the root to a JSON object. Paths may be encoded in
	JSONPath~\cite{Friesen2019}, a straightforward path language for JSON.
	
	We next introduce the schema for our relational encoding. We reserve attribute $O.\text{id}$ for an (internal) object identifier. The object identifier must be unique, but we do not impose any constraints on its semantics. In the following, we will simply use the line of code in the file containing the JSON Schema declaration (after pretty printing), where the scope of the object is first entered.
	
	We identify the labels of all properties reachable by a given path~$p$:
	\begin{multline*}
		L_p = \{l_i \mid \mbox{object} \{l_1: J_1, \dots, l_n: J_n\}
		\mbox{ is reachable by path } p\}
	\end{multline*}
	
	For each property label~$l_i$ in~$L_p$ where $J_i$ occurs as a basic value, we define a relational attribute $l_i.\text{value}$ that captures the basic value~$J_i$. These properties are considered to be candidates for tags.
	
	Further, for each property label~$l_i$ in~$L_p$, we  define an attribute $l_i.\text{type}$, capturing the subschema directly derived from its value~$J_i$.

	For each object reachable by path~$p$, we then insert one tuple into this relation, choosing
	some unique object identifier for each object. The attribute values are populated with the semantics described above;
	\emph{null} values mark missing entries.

	\begin{example}
		\label{ex:explain_relational_encoding}
		Table~\ref{fig:geojson_table} shows the encoding for the JSON objects from Figure~\ref{fig:geojson_data} in the array starting from line~3, reachable by the JSONPath
		\verb!/geometries[*]!. By~t1 and~t2, we abbreviate the subschemas directly derived from the JSON values, requiring an array of numbers (t1) and an array of arrays of numbers (t2), namely
		\begin{equation}\tag{t1}
			\begin{split}
				\texttt{\{ "type" }&\texttt{: "array",}\\
				\texttt{  "items"}&\texttt{: \{ "type":"number" \}\}}
			\end{split}
		\end{equation}
		and further
		\begin{equation}\tag{t2}
			\begin{split}
				&\texttt{\{ "type": }\texttt{"array", "items":\{}\\
				&\qquad\texttt{"type": "array",}\\ 
				&\qquad\texttt{"items": \{ "type": "number" \}\}\}} 
				\label{equ:array_of_arrays_of_int}
			\end{split}
		\end{equation}
	\end{example}

	\begin{table}[htb]
		\caption{Relational encoding for the objects in Figure~\ref{fig:geojson_data} that are reachable by path {\tt /geometries[*]} (JSONPath syntax). 
			Subschemas~{\rm t1} and~{\rm t2} abbreviated as in Example~\ref{ex:explain_relational_encoding}.}
		
		\label{fig:geojson_encoding}
		\label{fig:geojson_table}
		
		\small 
		\centering
		
		\begin{tabular}{r lll}
			\toprule
			$O$.id & {\tt type}.value & {\tt type}.type & {\tt coordinates}.type\\
			\midrule
			3 & Point & string & {\rm t1}
			\\
			5 & Point & string & {\rm t1}
			\\
			7 & LineString & string & {\rm t2}
			\\
			9 & LineString & string & {\rm t2}
			\\
			\bottomrule
		\end{tabular}

	\end{table}
	
	\paragraph{Dependencies}
	We next introduce dependencies over this relational encoding.
	Traditional functional dependencies (FDs) capture constraints that hold on all tuples of a relation.
	Moreover, conditional functional dependencies (CFDs)~\cite{DBLP:conf/icde/BohannonFGJK07}  are functional dependencies that hold on only a subset of the tuples. 
	While in their full generality, CFDs are a generalization of classical functional dependencies, we will focus on a very restricted subclass that is related to association rules~\cite{DBLP:conf/pkdd/RammelaereG18}, and that can be defined quite compactly.
	
	\begin{definition}
		If $\mathcal{A}$ is a set of attributes, then a \emph{unary constant conditional functional dependency over $\mathcal{A}$} (ucCFD) is an expression of the form
		$$
		[A=a] \rightarrow [B=b]
		$$
		where $A, B$ are attributes in~$\mathcal{A}$ and $a, b$ are constants from the domains of~$A$ and~$B$ respectively.
		A relation $R$ over $\mathcal{A}$ satisfies
		$[A=a] \rightarrow [B=b]$ if for each pair of tuples $s, t \in R$, $\pi_A(s) = \pi_A(t) = a$ implies $\pi_B(s) = \pi_B(t) = b$.
	\end{definition}

	\begin{example}
		The dependency below holds in Table~\ref{fig:geojson_table} and reads as follows:
		$$
		[\texttt{type}.\text{value}=\text{"Point"}] \rightarrow  
		[\texttt{coordinates}.\text{type} = \text{t1}]
		$$
		
		The left-hand-side, in brackets, declares a condition that must be satisfied for the dependency to hold: We consider all attributes where the value of attribute {\tt type}.value is the string \enquote{Point}.
		For all tuples where this condition is satisfied, the value of attribute {\tt coordinates}.type must be the subschema abbreviated as~t1. 
		
	\end{example}

	In our domain of application, namely JSON values,
	these dependencies express powerful constraints between property values and subschemas: 
	In the example dependency above, it is implied that if the value of property {\tt type} is the string constant \enquote{Point}, then the value of the sibling property {\tt coordinates} must conform to the subschema~t1.

	In the remainder of this article, we exclusively focus on the discovery of such 
	\emph{value-type} constraints, 
	where the attribute on the left-hand-side of a ucCFD is of the form \enquote{$A$.value} (the value of property~$A$, which we consider to be a candidate for a tag in a tagged union),
	and the attribute on the right-hand-side is of the form \enquote{$B$.type} (the subschema of the sibling property~$B$).
	
	Traditionally, tagged unions are declared by switching on the value of a \emph{single} tag property. This is also the recommended practice in JSON Schema~\cite{jsifthenelse}, and in agreement with what we observe in real-world data~\cite{DBLP:conf/er/BaaziziCGSS21}.
	We are therefore confident that our restriction to unary dependencies is justified.

	\section{Related Work}
	\label{sec:related_work}
	
	Our article builds upon the rich body of related work in the area of schema extraction and the theory of data dependencies.
	
	\sloppy 
	\paragraph{Schema and constraint definition.}
	Schema languages for semi-structured data are well-researched.
	
	\fussy
	XML was developed as a semi-struc\-tu\-red data format with implicit structural information and an optional explicit schema. The simplest schema language for XML is DTD (Document Type Definition). The lack of means to express data types and exact cardinalities in DTDs motivated the development of further schema languages, such as XML~Schema~\cite{xml-schema}, Schematron~\cite{schematron}, and RelaxNG~\cite{relaxng}. 
	All three support the definition of constraints. In XML Schema, alternatives can be defined by specifying conditions on path expressions (XPath). 
	In version~1.1 of XML Schema, the concept of \webref{https://www.w3.org/TR/xmlschema11-1/\#cAssertions}{assertions} allows to encode constructs such as tagged unions.
	Schematron can define rules with context information (XPath) and messages (that are sent in the success or error case). RelaxNG enables the definition of inference rules. This historical excursion shows the necessity of exact and expressive schema languages. The same holds true for the description of JSON data. The preliminaries on JSON Schema were already covered in \autoref{sec:jsonschema}. 
	
	\paragraph{Schema extraction.}
	
	In \textit{relational databases} which follow a schema-first approach, all available databases have an explicit schema stored in the databases catalog. 
	With semi-structured data, on the other hand, there are many datasets available that have been
	published without such explicit schema information.  Schema extraction (also known as reverse engineering) is therefore an important subtask in data profiling for semi-structured data. 
	
	\paragraph{XML schema extraction.} For schema extraction from XML documents, different approaches have been developed, (e.g.~\cite{garofalakis2000xtract, moh2000dtd,DBLP:conf/icde/HegewaldNW06,DBLP:journals/cj/KlempaKMSSSVNH15,10.1145/544220.544288,mlynkova_necasky_2013}). In all algorithms, a simple schema consisting of element and attribute names, nesting, optional and required information is derived.
	We are not aware of any approaches to extracting complex schema constraints, such as tagged unions.
	
	\paragraph{JSON schema extraction.}
	Early work on schema extraction from JSON data~\cite{DBLP:conf/btw/KlettkeSS15} adds --- besides the schema itself --- also the extraction of statistics and a detection of outliers. In~\cite{klettke2017uncovering}, the extraction of schema versions over time, as well as evolution operations mapping between consecutive schema versions, is presented. 
	
	Recent surveys of different schema extraction approach\-es were provided by \cite{DBLP:conf/er/ContosS20} (qualitative comparison) and \cite{enase22} (quantitative  comparison). Several of the examined approaches  also support the extraction of union types:  \cite{DBLP:conf/btw/KlettkeSS15} and \cite{DBLP:conf/iri/FrozzaMC18} use the JSON Schema keyword \texttt{oneOf}, while \cite{DBLP:journals/vldb/BaaziziCGS19}
	use the union type constructor of their own proprietary schema language. The authors of \cite{DBLP:conf/er/RuizMM15} encode the extracted schema in the XML Schema language, and encode union types using entity versioning, while \cite{DBLP:conf/icwe/IzquierdoC13} pursue an alternative approach of reducing different types to their most generic type. 
	
	But neither Klettke et al.~\cite{DBLP:conf/btw/KlettkeSS15} nor Frozza et al.~\cite{DBLP:conf/iri/FrozzaMC18} support tagged unions
	within their extraction of union types. The approach by Baazizi et al.~\cite{DBLP:journals/vldb/BaaziziCGS19} is based on type inference. They achieve scalability by inferring types in a MapReduce-based approach. The authors do discuss the challenge of extracting tagged unions, and describe an extension to their algorithm to address this challenge. However, there is no implementation or evaluation of this feature, or of any heuristics to prevent overfitting in tagged unions.
	
	In a further, recent contribution, Spoth et al.~\cite{DBLP:conf/sigmod/SpothKLHL21} focus on resolving ambiguities during schema extraction, such as sets encoded as objects rather than arrays. As their approach does not consider property values, it cannot detect tagged unions. 
	
	Durner et al.~\cite{DBLP:conf/sigmod/DurnerL021} recently presented an approach for fast analytics on semi-structured data. Using various algorithms, the JSON data is divided into {\em tiles}, and local schemas are extracted. However, tagged unions are neither considered nor recognized.

	A completely different approach for schema inference has been suggested in 
	\cite{DBLP:journals/is/GallinucciGR18}. In this work, a supervised learning method (based on the well-known C4.5 classification algorithm) is used for detecting hidden rules in the different variants of datasets. These rules can be either structure-based or value-based. An interesting observation of this work was that based on an empirical study (interviews), value-based rules are considered more important by human consumers than structure-based rules to distinguish variants.  
	The result of the approach is a decision tree that distinguishes the different schema variants with value-based or structure-based conditions on each edge. Even if the approach in the paper is very different from ours, we recognize the motivation that considering value-based conditions in heterogeneous databases is important.

	\paragraph{Dependencies.}
	For an overview of data dependencies, we refer to a comprehensive survey~\cite{9302878}, 
	and focus on functional, inclusion, and conditional functional dependencies in the following.
	
	\paragraph{Functional dependencies and inclusion dependencies.}

	Functional dependencies \cite{10.1007/3-540-45876-X_10, 10.1007/978-3-540-39403-7_27,DBLP:conf/icdt/KotW07,DBLP:journals/corr/Mior2021} and inclusion dependencies on semi-structured data  \cite{DBLP:conf/btw/KrusePN15,DBLP:journals/corr/Mior2021} define semantic constraints on data to guarantee certain data characteristics, to normalize data \cite{6719198}, and to ensure data quality. These constraints can also be applied in data cleaning. In \cite{DBLP:conf/sigmod/ChuIKW16}, an overview over different methods and the semantic constraints, rules, or patterns to detect errors in the data cleaning process is given. Schelter et al.~\cite{schelter2018automating} use declarative rules in unit tests to check the data quality (defined with several metrics). Some examples are: completeness of datasets, range conditions, certain cardinalities and constraints on statistics. Both approaches do not infer the semantic constraints from data but show how declarative rules can be leveraged during data preprocessing.

	The discovery of valid uniqueness constraints, functional dependencies and inclusion dependencies is a well-studied field in relational databases~\cite{kivinen1995approximate,mannila1992complexity,yao2008mining,10.1145/2882903.2915203,bauckmann2007efficiently} and even more relevant in JSON data (or, more generally, NoSQL data) because often these semantic constraints are not predefined in the NoSQL databases. They are implicitly available in the data but for data cleaning tasks, this information is required in form of rules and constraints. Hence, development of algorithms to derive explicit semantic constraints from NoSQL data is of particular importance.

	Methods for deriving semantic constraints from data can build upon  the algorithms developed for relational databases. Additionally, the algorithms have to scale with large volumes of data, and be robust despite the heterogeneity of datasets (variety) and low data quality (outliers in datasets). 
	
	Arenas and Libkin define functional and inclusion 
	dependencies for XML~\cite{10.1145/974750.974757}, as a basis for schema normalization. 
	More recently, Mior~\cite{DBLP:journals/corr/Mior2021} targets the mining of  functional and inclusion dependencies from JSON data.
	Both formalisms do not allow to capture the conditional functional dependencies required in our context.
	
	However, Mior compares the performance of dependency discovery in a relational encoding of the JSON data (termed \enquote{static unrolling}) with a dynamic unrolling technique, showing that the latter has superior runtime performance.
	Mior focuses on the discovery of functional dependencies and inclusion dependencies. While his approach does not consider the special case of conditional functional dependencies, which is relevant in our context, it is capable of deriving approximate dependencies, thus being more robust w.r.t.\  vagueness in the input data. 
	
	Kruse et al.\ developed in \cite{DBLP:conf/btw/KrusePN15} the 
	scalable discovery of inclusion dependencies from data. Scalability of the approach is achieved by two reasons: the approach is only concentrating on discovery of unary inclusion dependencies (consisting of one attribute on the left-hand and on the right-hand side) and is developing a distributed approach. Combining both, a very efficient algorithm can be developed for analyzing large datasets. The heterogeneity of datasets has not been considered in this approach. In previous work on inferring inclusion dependencies from JSON data~\cite{klettke2017uncovering}, we suggested an efficient method  that uses lattice characteristics to optimize the algorithm. To consider outliers, a threshold is introduced. With it, it is possible to derive related inclusion dependencies which are violated by a small number of outliers.   
	
	A further relaxation of functional dependency discovery has been suggested in \cite{DBLP:conf/er/HaiQW19} for data lakes. Here, outliers are not ignored, but  properties that have different yet similar labels  are combined. 
	This method can be used for data exploration and profiling of datasets with lower data quality.

	\paragraph{Conditional functional dependencies.}
	Conditional functional dependencies~\cite{DBLP:conf/icde/BohannonFGJK07} were first introduced for relational data, in the context of data cleaning. We apply conditional functional dependencies in a new context, the relational encoding of JSON objects. 
	
	In \cite{4812508,DBLP:journals/tkde/FanGLX11}, three  algorithms for CFD mining are proposed and evaluated: CTANE and FastCFD build upon existing algorithms for FD discovery, TANE and FastFD, respectively and are intended for general CFD discovery. The third algorithm is custom-designed for constant CFDs, as also targeted by us. 
	
	In~\cite{DBLP:journals/cj/LiLTY13}, additional rules for pruning the search space and consequently speeding up the mining process are proposed for constant CFDs. Further approaches, based on FD discovery and pattern mining, are explored in~\cite{DBLP:conf/pkdd/RammelaereG18}.

	\section{Approach}
	
	\label{sec:approach}
	
\begin{figure*}[ht]
		\centering

\definecolor{unusedclr}{RGB}{228,26,28} 
\definecolor{discoveryclr}{RGB}{55,126,184} 
\definecolor{extractionclr}{RGB}{77,175,74} 

\begin{tikzpicture}[
    scale=1,
    root/.style={draw, ellipse, minimum width=0.5cm, minimum height=0.4cm},
    treenode/.style={draw, rectangle, inner sep=0, rounded corners, minimum width=0.5cm, minimum height=0.35cm},
    leaf/.style={draw, rotate=90, rectangle, rounded corners=3pt, minimum width=0.4cm, minimum height=0.15cm},
    arrow/.style={-{Stealth}},
    cell1/.style={rectangle, black, inner sep=0, minimum width=0.27cm},
    cell2/.style={rectangle, black, inner sep=0, minimum width=1.9cm, minimum height=0.45cm},
    table/.style={draw, inner sep=0, row sep=-\pgflinewidth,column sep=-\pgflinewidth, matrix of nodes, text depth=0.5ex,text height=2ex},
    label/.style={align=center, font=\bfseries\small, text depth=0pt, anchor=base},
    number/.style={draw, fill=black!10, circle, minimum size=0.35cm, inner sep=0, text=black},
    result/.style={font=\small, align=left}
]

    \newcommand{\corner}{0.3}

    \tikzset{
        file-small/.pic = {
            \node[rectangle, minimum height=1cm, minimum width=1.1cm] (f) {};
            \draw[fill=white] (f.south west) -- (f.north west) -- ($(f.north east)+(-\corner,0)$) coordinate (ctl)
            -- ($(f.north east)+(0,-\corner)$) coordinate (cbr) |- cycle;
            \node[align=left,font=\small, anchor=west] at (f.west) {\tikzpictext};
            \fill (ctl) |- (cbr) -- cycle;
        },
        file-narrow/.pic = {
            \node[rectangle, minimum height=1cm, minimum width=0.8cm] (f) {};
            \draw[fill=white] (f.south west) -- (f.north west) -- ($(f.north east)+(-\corner,0)$) coordinate (ctl)
            -- ($(f.north east)+(0,-\corner)$) coordinate (cbr) |- cycle;
            \node[align=left,font=\small, anchor=west] at (f.west) {\tikzpictext};
            \fill (ctl) |- (cbr) -- cycle;
        },
        file-large/.pic = {
            \node[rectangle, minimum height=1.2cm, minimum width=1.6cm] (f) {};
            \draw[fill=white] (f.south west) -- (f.north west) -- ($(f.north east)+(-\corner,0)$) coordinate (ctl)
                -- ($(f.north east)+(0,-\corner)$) coordinate (cbr) |- cycle;
            \node[align=left,font=\small, anchor=west] at (f.west) {\tikzpictext};
            \fill (ctl) |- (cbr) -- cycle;
        }
    }

    \pic[pic text={JSON\\Schema}, local bounding box=schema] at (0,-0.2) {file-small};

    \pic[local bounding box=json-bg] at (-0.1, -1.8) {file-narrow};
    \pic[pic text={JSON}, local bounding box=json] at (0, -1.9) {file-narrow};

    \pic[pic text={JSON\\Schema\\ \{allOf:[$S$,$T$]\}}, local bounding box=ite] at (13.2, -0.2) {file-large};

    \node[root] (pt-root) at (2.95,-0.7) {};

    \node[treenode] (pt-11) at ($(pt-root)+(-0.4, -0.7)$) {};
    \node[treenode] (pt-12) at ($(pt-root)+(0.4, -0.7)$) {};
    \draw[arrow] (pt-root) -- ++(0,-0.3) -| (pt-11.north);
    \draw[arrow] (pt-root) -- ++(0,-0.3) -| (pt-12.north);

    \foreach \x/\l in {-0.8/A, 0/B, 0.8/C} {
        \node[treenode] (pt-\l) at ($(pt-11)+(\x, -0.7)$) {\l};
        \draw[arrow] (pt-11) -- ++(0,-0.3) -| (pt-\l.north);

        \foreach \y/\i in {-0.26/1, 0/2, 0.26/3} {
            \node[leaf] (pt-\l\i) at ($(pt-\l)+(\y, -0.75)$) {\footnotesize...};
            \draw[arrow] (pt-\l) -- ++(0,-0.3) -| (pt-\l\i.east);
        }
    }

    \draw[red] ($(pt-A.south west)+(-0.05,-0.05)$) rectangle ($(pt-C.north east)+(0.05,0.05)$);

    \node[label, left=0.3 of pt-root] {Parse\\Tree};

    \matrix[table, row 1/.style={nodes={cell1}}, nodes={inner sep=0, minimum height=0.2cm}]
    (comp-recs) at ($(pt-C.east)+(1.15,0.1)$) {
        X & Y & Z \\\hline
        $\cdot$ & $\cdot$ & $\cdot$ \\
        $\cdot$ & $\cdot$ & $\cdot$ \\
        $\cdot$ & $\cdot$ & $\cdot$ \\
    };
    \node[label, above=0.1 of comp-recs] (comp-recs-label) {Comp.\\Records of Rel.\\Encoding};

    \matrix[matrix of nodes, draw, column sep=0, nodes={inner sep=0, minimum height=0.33cm}]
    (list-indices) at ($(comp-recs.east)+(1.45,0)$) {
        $\mathrm{\pi_{\mathsf{X}}}$ & $=$ & $\left\{\dots\right\}$\\
        $\mathrm{\pi_\mathsf{Y}}$ & $=$ & $\left\{\dots\right\}$ \\
        $\mathrm{\pi_{\mathsf{Z}}}$ & $=$ & $\left\{\dots\right\}$\\
    };
    \node[label, anchor=base] (list-indices-label)
        at (list-indices |- comp-recs-label.base) {Position\\List\\Indices};

    \matrix[table, nodes={cell2}]
    (cand-list) at ($(list-indices.east)+(1.6,0)$) {
        $\scriptstyle\mathrm{[\mathsf{X}=v_1] \rightarrow [\mathsf{Y}=t_1]}$ \\\hline
        $\scriptstyle\mathrm{[\mathsf{X}=v_2] \rightarrow [\mathsf{Y}=t_2]}$ \\\hline
        $\scriptstyle\mathrm{[\mathsf{X}=v_3] \rightarrow [\mathsf{Z}=t_3]}$ \\
    };
    \node[label, anchor=base, above=0.1 of cand-list] {CFDs\\Candidate List};

    \matrix[table, nodes={cell2}] (fd-id-list)
    at ($(cand-list.east)+(1.6,0)$) {
        $\scriptstyle\mathrm{[\mathsf{X}=v_1]\rightarrow[\mathsf{Y}=t_1]}$ \\\hline
        $\scriptstyle\mathrm{[\mathsf{X}=v_2] \rightarrow [\mathsf{Y}=t_2]}$ \\\hline
        \color{red}\sout{$\scriptstyle\mathrm{[\mathsf{X}=v_3] \rightarrow [\mathsf{Z}=t_3]}$} \\
    };
    \node[label, above=0.1 of fd-id-list, xshift=1.15cm] {CFDs List};

    \coordinate (ah1) at (schema.20);
    \node[number] (eight) at ($(ite.west |- ah1)+(-0.45,0)$) {8};

    \draw[arrow, font=\small] (schema.east |- ah1) -- (eight) node[result, above, pos=0.13] {Schema $S$};
    \draw[arrow] (eight) -- (ite.west |- ah1);

    \coordinate (ah2) at ($(json)+(0,-0.15)$);
    \draw[arrow] (json.north -| schema.south) -- (schema) node[midway, right, number, xshift=0.1cm] {1};
    \draw[arrow] (json.east |- ah2) -- ($(pt-A.west |- ah2)+(-0.05,0)$) node[pos=0.4, number] {2};
    \draw[arrow] ($(pt-C.east |- ah2)+(0.05,0)$) -- (comp-recs.west |- ah2) node[pos=0.45, number] {3};
    \draw[arrow] (comp-recs.east |- ah2) -- (list-indices.west |- ah2) node[pos=0.38, number] {4};
    \draw[arrow] (list-indices.east |- ah2) -- (cand-list.west |- ah2) node[pos=0.38, number] {5};
    \draw[arrow] (cand-list.east |- ah2) -- (fd-id-list.west |- ah2) node[pos=0.4, number] {6};
    \draw[arrow] (fd-id-list.north -| eight) -- (eight) node[pos=0.6, number] {7};

    \node[result] at ($(ite.south)+(-2.15,-0.0)$) {if-then-else\\Constraints $T$};

    \coordinate (discovery-mid) at ($(comp-recs.west)!.5!(cand-list.east)$);
    \node[align=center, font=\small] (extraction)
        at ($(json-bg)+(0.35,-1)$)
        {Schema Extrac-\\tion, e.g.~\cite{DBLP:conf/btw/KlettkeSS15}};
    \node[anchor=base, font=\small] (discovery) at ($(discovery-mid |- extraction.base)$)
        {Dependency Discovery};

    {[on background layer]
        \fill[extractionclr!25] ($(schema.north -| extraction.west)+(0,0.2)$) rectangle ($(extraction.south east)$);
        \fill[discoveryclr!25] ($(comp-recs.north west)+(-0.25,0.12)$) rectangle ($(discovery.south -| cand-list.east)+(0.25,0)$);
    }

\end{tikzpicture}
		\caption{System architecture overview. (1) A third-party tool is used to extract a JSON Schema description~$S$ of JSON input data. Steps~(2) through~(7) visualize the discovery of tagged unions as schema~$T$, and are described in Section~\ref{sec:approach}. In step~(8), the schemas~$S$ and~$T$ are composed into a composite schema. Boxed areas capture state-of-the-art algorithms integrated in our architecture. A, B and C are JSON property labels while X, Y and Z are attributes in our relation encoding (e.g., A.value).}
		\label{fig:arch}
	\end{figure*}
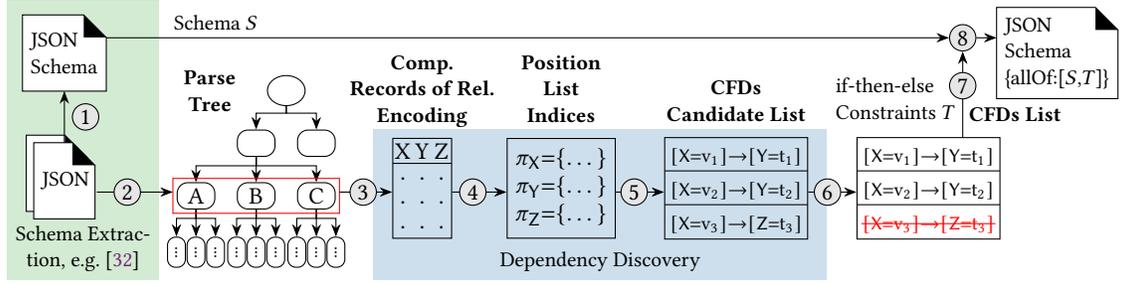

	Our end-to-end approach is sketched in Figure~\ref{fig:arch}. In our upcoming walk-through of the architecture, we first focus on the basic approach, 
	as well as our heuristics against overfitting and schema bloat. 
	We then outline future work extensions to robustly generalize our approach to a larger family of tagged unions.

	\subsection{Architecture}
	
	\label{sec:arch}
	
	\paragraph{Composability.} We design our approach to be composable with existing schema extraction algorithms:
	Given a collection of JSON values, we extract a JSON Schema description~$S$, using any third-party tool.
	Along the process sketched along the bottom, we further derive the declaration of tagged unions, denoted~$T$,
	a sequence of (possibly nested) JSON Schema
	{\tt if-then-else} statements.

	By the conjunction of the subschemas as \\
	\verb!{"allOf": [!S\verb!,!$T$\verb!]}!,
	the composite schema requires the JSON values to satisfy both subschemas, imposing  tagged unions on  schema~$S$. This approach  allows us to focus on the particular challenge of identifying tagged unions, while conveniently leveraging state-of-the-art schema extraction tools for describing the other aspects of the schema. We leave it to future work to recursively merge the composite schema, thereby improving the readability as well as the locality of the interacting constraints.

	\paragraph{Relational encoding.} In a first step towards the discovery of tagged unions, we parse all JSON values into a single parse tree (adding a virtual root node), and determine the set of all labeled paths from the  root to a JSON object. 
	For each such path, we then derive the relational encoding of all properties in the reachable objects. 
	We have already introduced this data structure in Section~\ref{sec:preliminiaries}, and resume the discussion of our running example.

	\begin{example}
		Consider the GeoJSON value in Figure~\ref{fig:geojson_data}. 
		Table~\ref{fig:geojson_encoding} shows the relational encoding of the four objects 
		reachable by the path {\tt /geometries[*]}. The role of the unique object identifier is to preserve cardinalities in repeating attribute values,
		allowing us to later filter dependencies based on their support in the input.
	\end{example}
	
	The running example has been designed to be simple, and it ignores challenges such as  properties with varying types, dealing with several properties whose values have basic types, or
	null-values in the relational encoding. Our approach can be extended to robustly handle these cases as well, and we refer to Section~\ref{sec:extensions} for a discussion on the most interesting future generalizations.

	\paragraph{Dependency discovery.}
	
	Based on the relational encoding, we discover conditional functional dependencies.
	We restrict ourselves to unary constant conditional dependencies in our relational encoding, of the form
	$$
	[A.\text{value}=c] \rightarrow [B.\text{type} = \sigma]
	$$
	where the left-hand-side denotes the value of the candidate tag,
	and the right-hand-side a property with an implied subschema.
	Above, $A$ and $B$ are distinct property labels, $c$ is a basic-value constant,
	and~$\sigma$ denotes the implied subschema. 
	Recall that with the dot-notation, we distinguish the value of JSON property~$A$ (i.e., $A.$value) from the subschema of sibling property~$B$ (i.e., $B$.type) in our relational encoding of JSON objects.

	\begin{example} \label{ex:fds_for_geojson}
		From the relational encoding in Table~\ref{fig:geojson_table}, we derive the following
		dependencies, again using the abbreviations~t1 and~t2 for the subschemas introduced earlier:
		{\small
			\begin{align*}
				[\mathtt{type}.\text{value}=\text{"Point"}] &\rightarrow [\mathtt{coordinates}.\text{type} = \text{t1}] \\ 
				[\mathtt{type}.\text{value}=\text{"LineString"}] &\rightarrow [\mathtt{coordinates}.\text{type} = \text{t2}] 
			\end{align*}
		}
		
		Recall that in our application context, they are interpreted as follows.
		Given a JSON object reachable by the path \verb!/geometries[*]!,
		if the value of property {\tt type} is \enquote{Point},
		the value of the sibling property \texttt{coordi\-nates} must conform to the subschema~t1 (an array of numbers). If the value of property {\tt type} is \enquote{LineString}, then the value of sibling property \texttt{coordi\-nates} must conform to the subschema~t2 (an array of points).
	\end{example}

	In the discovery of these dependencies, we leverage state-of-the-art algorithms~\cite{DBLP:books/acm/IlyasC19,10.1145/2882903.2915203,DBLP:conf/edbt/SchirmerP0NHMN19,8731407} (originally developed for the discovery of traditional functional dependencies),
	such as a compressed-records encoding of the relation, computing position list indexes (PLIs) and producing a sequence of candidate dependencies. PLIs are calculated for each column of the relational encoding and allow to effectively group all rows (identified by their row number) with the same value in the same subset.
	
	In the following, we outline the general idea.
	Inferring conditional functional dependencies from PLIs can be achieved by finding inclusions between them: given PLIs $\pi_\textit{A.value} = \{\{1,2,3\},\{4,5,6\}\}$ of attribute $A.$value
	and $\pi_\textit{B.type} = \{\{1,2,3,4\},\{5\},\{6\}\}$ of attribute $B.$type, we receive the inclusion $\{1,2,3\} \subset \{1,2,3,4\}$, allowing us to infer that the values of JSON property A corresponding to rows 1, 2 and 3 determine the subschemas of the sibling property B. 
	
	More sophisticated algorithms developed for the discovery of conditional functional dependencies (CFDs), such as CFDMiner, CTANE and FastCFD~\cite{4812508}, could be deployed just as well. As our current restrictions, most notably the one to unary dependencies, obfuscate many of the challenges in FD and CFD discovery, we opted for our simpler approach.

	\paragraph{Pruning candidate dependencies.}
	
	To avoid overfitting in the derived schemas, we heuristically prune candidate dependencies, as described in the upcoming Section~\ref{sec:heuristics}.

	\paragraph{Finalization.} In a final step, the conditional functional dependencies are transformed into nested {\tt if-then-else} statements in JSON Schema~$T$. This step is merely technical, as is the construction of the composite schema.

	\subsection{Heuristics}
	\label{sec:heuristics}

	In our approach, an inherent challenge is dealing with high numbers of discovered dependencies, which may lead to overfitting the derived schema w.r.t.\ the input data.
	We next  distinguish default heuristics and heuristics that are configurable by an expert user.

	\subsubsection{Default Heuristics}
	\label{sec:default_heuristics}
	
	The following heuristics are always applied, as they prevent schema bloat due to overfitting.
	
	\paragraph{Single-valued attributes.}
	
	In discovering dependencies, we ignore all attributes from the relational encoding with a single-valued domain. In Table~\ref{fig:geojson_table}, this concerns the attribute {\tt type}.type.

	\paragraph{Unique attributes.}
	Unique property values are recognized as conditional functional dependencies, and ultimately, cause overfitting and schema bloat. 
	In dependency discovery, we therefore ignore all attributes where the domain consists of unique values (such as the object identifier $O$.id, which does not appear in the input data, and need not be considered in dependency discovery).

	\paragraph{Union rule.} 
	We apply the union rule to the discovered dependencies, so that one tag may imply the subschemas for several sibling properties. 
	This improves schema succinctness.
	
	\subsubsection{Configurable Heuristics}
	\label{sec:threshold}
	The following heuristics is reasonable in many cases. As it is not universally applicable, we  make it configurable.
	\paragraph{Minimum Threshold $\pi_\textit{min}$.} 
	
	Especially in large data\-sets, it is to be expected that dependencies are inferred that have no real semantics, despite us ignoring attributes with unique values. 
	Our implementation has a configurable \emph{minimum threshold}: Dependencies with insufficient support in the input are then ignored during the generation of {\tt if-then-else} statements. 
	
	By allowing to configure this threshold, either as an absolute or relative value, the number of discovered dependencies can be influenced.
	Our experiments evaluate this effect.
	
	\subsection{Future Generalizations of the Approach}
	\label{sec:extensions}
	
	In GeoJSON data, which served as our motivation for the discovery of tagged unions, the geometry objects are highly regular. When inspecting further instances of tagged unions in real-world data, we encounter variations that require us to generalize our approach. For illustration, we again resort to an example.

	\begin{figure}[thb]
		\centering
		
		\begin{subfigure}{\linewidth}
			\centering
			\begin{queryboxlink}{}{}
				\begin{lstlisting}[style=query]
					[ { "condition": "minecraft:time_check",
						"period": 24000,
						"value": { "min": 0.0 } } ]
				\end{lstlisting}
			\end{queryboxlink}
			\caption{\webref{https://github.com/bafomdad/uniquecrops/blob/96697872796ca0b6700442116f98eaadd221c8cb/src/main/resources/data/uniquecrops/loot_tables/blocks/crop_dyeius.json}{The value of property {\tt min} (line~3) is numeric}.}
		\end{subfigure}

		\vspace{0.2cm}
		
		\begin{subfigure}{\linewidth}
			\begin{queryboxlink}{}{}
				\begin{lstlisting}[style=query]
					[ { "condition": "minecraft:weather_check",
						"raining": false,
						"thundering": false },
					{ "condition": "minecraft:time_check",
						"period": 24000,
						"value": {
							"min": {
								"target": {
									"name": "%StartSunburn",
									"type": "minecraft:fixed" },
								"score": "Daytime",
								"type": "minecraft:score" } } } ]
				\end{lstlisting}
			\end{queryboxlink}
			\caption{\webref{https://github.com/IceFreez3r/OneHeart/blob/b2022fb00f08b150e8b170d8f0b403ee9aa224c7/data/oneheart/predicates/sunburn.json}{The value of property {\tt min} (line~7)  is of type object.}}
		\end{subfigure}
		
		\caption{Minecraft JSON data (snippets edited and shortened, clickable links to GitHub source in the subcaptions). Property {\tt condition} functions as the tag in the tagged union.}
		\label{fig:minecraft_values}
	\end{figure}

	\begin{example} \label{ex:minecraft}
		We consider JSON data used in the Minecraft game, as shown in Figure~\ref{fig:minecraft_values}.
		The property {\tt condition} functions as a tag, and distinguishes notions such as \enquote{time checks} and \enquote{weather checks}, among others. 
		Depending on the tag value, different sibling properties exist: Given that the tag value is a \enquote{weather check}, sibling properties labeled {\tt raining} and {\tt thundering} exist.
		Given that the tag value is a 
		\enquote{time check}, sibling properties labeled {\tt period} and {\tt value} exist. 
	\end{example}
	
	To extend our approach to such more general forms of tagged unions, we need to be able to detect conditional dependencies between the value of a candidate tag, and the existence of a sibling property (which we term \emph{value-label} constraints, to distinguish from the value-type constraints that we already handle). 
	Here, a specific challenge is to handle optional sibling properties, which are not always present. 
	
	\begin{example}
		Resuming the previous example of a Minecraft \enquote{time check}, the  subschema for property {\tt value} is an object with a property {\tt min}.
		The subschema of property {\tt min}, however, may vary. In the top example, it is a numeric value. 
		In the bottom example, it is a nested object.
		Our approach outlined so far cannot detect this conditional functional dependency, because the subschemas of the sibling property differ.
	\end{example}
	
	To detect value-type dependencies where the types on the right-hand-side may vary, we need to relax the recognized subschema. In our current prototype implementation, we introduce attributes {\tt min}.type$_1$, \dots, {\tt min}.type$_k$ (e.g., for a fixed limit $k=6$), where we describe the subschema with a varying degree of detail. For instance, from a shallow {\tt min}.type$_1$= {\tt \{"type": "object\}} to a very fine-grained description that fully captures all nested objects. In pruning candidate dependencies, we only keep the most detailed subschema for which a CFD still holds. This allows us to detect a common subschema in many practical cases.

	\section{Experiments}
	\label{sec:experiments} 
	\begin{table*}[ht!]
	\caption{Summary of experimental results. Links embedded in the PDF refer to the original dataset ($D$). We state the dataset format (TopoJSON or GeoJSON) and
		report its size ($|D|$) in lines of code (LoC), the size of the conventionally extracted schema~$S$ (in LoC) and the number of theoretically detectable Geo-/TopoJSON CFDs ($Det_{Geo}$). For different threshold settings~$\pi_\textit{min}$, we produce a subschema~$T$ encoding the tagged unions. \enquote{Ratio$T$} reports the share of~$T$ w.r.t.\ the size of  the composite schema. Further, the number of CFDs initially discovered (w/o heuristics), after applying threshold~$\pi_\textit{min}$ (see Section~\ref{sec:threshold}), and after applying the remaining default heuristics (see Section~\ref{sec:default_heuristics}) are reported. 
		The last column reports the result of checking whether dataset~$D$ is valid w.r.t.\ the extracted composite schema (a checkmark symbol~\enquote{$\checkmark$} denotes a successful check).
		The checkmarks are clickable links, embedded in the PDF, to the composite schema.
	}
	\label{table:exp}
	
	\centering
	\begin{tabular}{@{}llrrrrrrrrrrc@{}} 
		\toprule
		\thead[l]{Dataset D} & \thead[l]{Format} & {\thead{|D| \\(LoC)}} & {\thead{|S|\\(LoC)}} & {\thead{$\textit{Det}_\textit{Geo}$}}  & \thead{$\pi_\textit{min}$} &  {\thead{|T|\\(LoC)}} & {\thead{Ratio$T$}} & {\thead{CFDs w/o \\Heuristics}} & {\thead{CFDs w/ \\ $\pi_\text{min}$}} & {\thead{CFDs w/ $\pi_\text{min}$\\ and Heuristics}} & \thead{Valid} \\
		\midrule
		& & & & & 50\% & 126 & 23.6\% & & 11 & 4 & \href{https://github.com/sdbs-uni-p/schema-inference-repro/blob/main/artifacts/original_results/heuristics_and_threshold_default/MapGermany/threshold-50/schema-ite.json}{\checkmark} \\ 
		\href{https://github.com/sdbs-uni-p/schema-inference-repro/blob/main/artifacts/input/wija/wija_Germany.json}{Map Germany} & T & 181 585 & 405 & 6 & 35\% & 158  &  27.9\% & 44 & 14 & 5 & \href{https://github.com/sdbs-uni-p/schema-inference-repro/blob/main/artifacts/original_results/heuristics_and_threshold_default/MapGermany/threshold-35/schema-ite.json}{\checkmark} \\
		& & & & &15\% & 190 & 31.7\% & & 20 & 6 & \href{https://github.com/sdbs-uni-p/schema-inference-repro/blob/main/artifacts/original_results/heuristics_and_threshold_default/MapGermany/threshold-15/schema-ite.json}{\checkmark} \\
		
		\rowcolor{gray!10}
		& & & & & 50\% & 95 & 20.7\% & & 6 & 3 & \href{https://github.com/sdbs-uni-p/schema-inference-repro/blob/main/artifacts/original_results/heuristics_and_threshold_default/MapEU/threshold-50/schema-ite.json}{\checkmark} \\ \rowcolor{gray!10}
		\href{https://github.com/sdbs-uni-p/schema-inference-repro/blob/main/artifacts/input/AtelierCartographie/AtelierCartographie_EU.json}{Map EU}& T & 122 380 & 359 & 6 & 35\% & 95 & 20.7\% & 10 & 6 & 3 & \href{https://github.com/sdbs-uni-p/schema-inference-repro/blob/main/artifacts/original_results/heuristics_and_threshold_default/MapEU/threshold-35/schema-ite.json}{\checkmark} \\\rowcolor{gray!10}
		& & & & & 15\% & 127 & 25.9\% & & 8 & 4 & \href{https://github.com/sdbs-uni-p/schema-inference-repro/blob/main/artifacts/original_results/heuristics_and_threshold_default/MapEU/threshold-15/schema-ite.json}{\checkmark} \\
		
		& & & & & 50\% & 150 & 10.0\% & & 76 & 4 & \href{https://github.com/sdbs-uni-p/schema-inference-repro/blob/main/artifacts/original_results/heuristics_and_threshold_default/WorldGenBerlin/threshold-50/schema-ite.json}{\checkmark} \\ 
		\href{https://github.com/sdbs-uni-p/schema-inference-repro/blob/main/artifacts/input/reinterpretcat/reinterpretcat_berlin.osm.json}{WorldGen Berlin} & G & 56 441 & 1 339 & 11 & 35\% & 182 & 11.9\% & 1 243 & 109 & 5 & \href{https://github.com/sdbs-uni-p/schema-inference-repro/blob/main/artifacts/original_results/heuristics_and_threshold_default/WorldGenBerlin/threshold-35/schema-ite.json}{\checkmark} \\
		
		& & & & & 15\% & 278 & 17.1\% & & 184 & 8 & \href{https://github.com/sdbs-uni-p/schema-inference-repro/blob/main/artifacts/original_results/heuristics_and_threshold_default/WorldGenBerlin/threshold-15/schema-ite.json}{\checkmark} \\
		
		\rowcolor{gray!10}
		& & & & & 50\% & 54 & 9.8\% & & 459 & 2 & \href{https://github.com/sdbs-uni-p/schema-inference-repro/blob/main/artifacts/original_results/heuristics_and_threshold_default/USHouseMembers/threshold-50/schema-ite.json}{\checkmark} \\ \rowcolor{gray!10}
		\href{https://github.com/sdbs-uni-p/schema-inference-repro/blob/main/artifacts/input/stenson/stenson_members.json}{US House Members} & G & 22 745 & 493 & 2 & 35\% & 54 & 9.8\% & 719 & 470 & 2 & \href{https://github.com/sdbs-uni-p/schema-inference-repro/blob/main/artifacts/original_results/heuristics_and_threshold_default/USHouseMembers/threshold-35/schema-ite.json}{\checkmark} \\ \rowcolor{gray!10}
		& & & & & 15\% & 54 & 9.8\% & & 660 & 2 & \href{https://github.com/sdbs-uni-p/schema-inference-repro/blob/main/artifacts/original_results/heuristics_and_threshold_default/USHouseMembers/threshold-15/schema-ite.json}{\checkmark} \\
		
		& & & & & 50\% & 38 & 20.5\% & & 18 & 1 & \href{https://github.com/sdbs-uni-p/schema-inference-repro/blob/main/artifacts/original_results/heuristics_and_threshold_default/WildlifeSites/threshold-50/schema-ite.json}{\checkmark} \\ 
		\href{https://github.com/sdbs-uni-p/schema-inference-repro/blob/main/artifacts/input/CalderdaleCouncil/Local%20Wildlife%20Sites.json}{Wildlife Sites} & G & 876 482 & 143 & 2 &35\% & 38 & 20.5\% & 360 & 18 & 1 & \href{https://github.com/sdbs-uni-p/schema-inference-repro/blob/main/artifacts/original_results/heuristics_and_threshold_default/WildlifeSites/threshold-35/schema-ite.json}{\checkmark} \\
		& & & & & 15\% & 92 & 38.5\% & & 50 & 3 & \href{https://github.com/sdbs-uni-p/schema-inference-repro/blob/main/artifacts/original_results/heuristics_and_threshold_default/WildlifeSites/threshold-15/schema-ite.json}{\checkmark} \\
		\bottomrule
	\end{tabular}
	
\end{table*}

	We next describe our experiments using real-world geo-spatial datasets encoded in JSON. For notational simplicity, we will write \enquote{CFDs} when we refer to the subfamily of unary constant conditional functional dependencies, as introduced in Section~\ref{sec:preliminiaries}. 
	
	\subsection{Setup}
	
	\paragraph{Implementation.}
	Our prototype is implemented in Python,
	using the library \href{https://github.com/c0fec0de/anytree}{anytree} for managing parse trees.
	For schema validation, we use the Python library \href{https://github.com/python-jsonschema/jsonschema}{jsonschema} (clickable links in the PDF).
	
	For the \enquote{third-party} schema extraction tool, we employ a tool that we have built in earlier work~\cite{DBLP:conf/btw/KlettkeSS15}. Note that our approach is designed to work with any other tool for JSON Schema extraction.
	Also, since we construct the composite schema by combining two subschemas, we are not restricted to Python-based tools.
	
	To confirm the composability of our approach, we have successfully built a variant of our architecture based on the schema extraction tool by Frozza et al.~\cite{DBLP:conf/iri/FrozzaMC18, frozza:schema_discovery} (with minor technical adaptions), but we do not report the results, since they do not contribute new insights.
	
	\paragraph{Artifact availability and reproducibility.}
	Our research artifacts (code, data, extracted schemas), as well as a fully automated reproduction package, are available online~\cite{zenodo:repro}. 
	Additionally, in the PDF version of this article, clickable links in Table~\ref{table:exp} allow our readers to directly inspect the input datasets, as well as the extracted schemas.
	
	\paragraph{Datasets.}
	We consider the five JSON datasets listed in Table~\ref{table:exp}, all in the GeoJSON or the related TopoJSON format. For simplicity, we will commonly treat the formats GeoJSON and TopoJSON synonymously, in discussing our results.

	We identified these datasets by manually searching open data servers, as well as by performing a search over all open-source licensed repositories on GitHub using Google BigQuery.
	Our selection criteria were (1)~a sufficiently large document size and (2)~that at least two 
	different types of geometries co-occur, so that we may be able to infer tagged unions.
	
	We briefly describe the datasets:
	\begin{compactitem}
		\item 
		\textit{Map Germany} is a TopoJSON dataset, describing a map of Germany and surrounding areas, including cities, urban areas, and rivers. This dataset uses geometry types Point, LineString, MultiLineString, Polygon and MultiPolygon.
		
		\item 
		\textit{Map EU} is a TopoJSON dataset. Its structure is comparable to \emph{Map Germany} but describes countries in the EU in detail and contains the borders of some adjacent countries.
		
		\item 
		\textit{WorldGen Berlin} is a GeoJSON dataset from OpenStreetMap.
		
		\item 
		\textit{US House Members} lists members of the US House of representatives, including their place of birth, hometown and congressional district, encoded in GeoJSON.
		
		\item 
		Wildlife Sites contains wildlife sites in West Yorkshire, described either as a GeoJSON Polygon or MultiPolygon.
		
	\end{compactitem}
	
	The datasets encode GeoJSON or the related TopoJSON format (distinguished as formats \enquote{T} and~\enquote{G} in the table).
	We list the size of each dataset in lines of code after pretty-printing (column~$|D|$).
	
	In lieu of a ground truth describing the number of conditional functional dependencies in the data that is not GeoJSON-encoded, we report in the table the number of detectable CFDs in the GeoJSON-parts only ($Det_{Geo}$). We manually determine this value for each dataset, inspecting all distinct types of GeoJSON objects (e.g., Point, LineString, Polygon) on each different path. We ignore all paths where the value of the GeoJSON {\tt type} does not vary. While these can be detected as CFDs from the data, they are not meaningful for recognizing tagged unions, and should be modeled, for instance, simply as constant values instead.
	
	\paragraph{Execution environment.}
	Our experiments were conducted 
	on an off-the-shelf notebook with an Intel i7-1165G7 CPU with 4.7GHz and 16GB of main memory.

	\subsection{Experimental Design}
	
	\paragraph{Workflow.}
	For each data collection, we extract composite schemas choosing three settings for the \emph{minimum threshold}~$\pi_\textit{min}$: an aggressive setting of  50\%, requiring a conditional functional dependency to occur in at least half of the tuples in the relational encoding of all objects reachable by a given path, and more lenient settings with 35\% and 15\%, respectively.
	
	For each threshold setting, we generate schemas~$S$ (from the third-party tool) and~$T$ (encoding the tagged unions), as well as the composite schema. 
	
	Each composite schema is validated against the specification of JSON Schema~\mbox{Draft-07}, confirming that the composite schema conforms. The
	input datasets are further validated against the extracted composite schema, checking  for any logical errors in schema extraction.

	\paragraph{Metrics.}
	In our quantitative assessment, we report schema sizes (after syntactic normalization by pretty-printing), the number of conditional functional dependencies initially discovered, and after filtering according to our heuristics.
	
	We use $\textit{Det}_\textit{Geo}$  as a target in discussing recall of dependency discovery in the GeoJSON part of the input.

	\subsection{Results}
	
	Table~\ref{table:exp} summarizes our results. 
	Note that all composite schemas successfully validate against Draft-07, and further, that all JSON datasets validate against the composite schemas.
	
	Decreasing the threshold~$\pi_\textit{min}$ leads to more conditional functional dependencies being discovered. The share of the subschema encoding tagged unions compared to the entire composite schema ranges between 10\% and up to approx. 38\%.
	While the 38\% share might appear to be large, the composite schema has less than 250 lines in total, compared to 800K lines of JSON input. Thus, this dataset is highly regular in its structure, and the extracted schema comparatively compact.
	
	For the bottom three datasets, we detect conditional functional dependencies in the hundreds, even more than one thousand in the case of the WorldGen~Berlin data.
	However, applying the threshold and the other heuristics drastically reduces the dependencies towards the number of tagged unions that we expect to find (c.f.\ the target~$\textit{Det}_\textit{Geo}$). Note that there are only six distinct geometries in GeoJSON which may, however, occur on different paths, leading to more than six \verb!if-then-else! statements in the GeoJSON part. 
	
	After applying heuristics and a threshold of 15\%, we only miss two detectable CFDs in Map~EU and three in WorldGen~Berlin. Missed CFDs can be attributed to the threshold being too coarse.
	
	Manually inspecting the final dependencies, we  find one unexpected dependency for Wildlife~Sites with a threshold of 15\%. This is the only false positive dependency across all datasets. It results in 
	a tagged union being declared for a specific date, which is not semantically meaningful. Further, this dependency is outside the GeoJSON-encoded part of the data, explaining why the number of detected CFDs is higher than $Det_{Geo}$.
	Thus, we have a single case of overfitting for the datasets analyzed.
	Overall, we observe very high precision in recognizing the GeoJSON/TopoJSON dependencies resident in the data.

	We do not focus on runtime evaluation, as our implementation is prototypical and unoptimized.
	Yet to provide a general perspective,
	we share our observation that runtimes vary greatly between datasets, ranging from roughly one second to approximately one minute. %
	This is a waiting time which we deem acceptable for a non-interactive, irregular task. 
	
	Since our approach is main-memory based, memory is a physical limitation to the inputs that we can process.
	For the datasets considered here, 16GB of RAM are sufficient to run our experiments.
	
	\subsection{Discussion}
	\label{sec:discussion}
	The comparison between the number of dependencies found with different heuristics shows the effectiveness of these heuristics in pruning dependencies. In particular, this concerns the configuration knob represented by \emph{minimum threshold}. 
	
	Our approach to setting this threshold is rather coarse, with 50\% obviously too rigid, but 15\% delivering meaningful results. In future work, we plan to investigate how to auto-adjust the threshold, based on statistical distributions obtained by profiling the input.
	
	An obvious threat to the generalizability of our results --- however promising they are --- is that the GeoJSON datasets are highly regular in their structure.  In the GeoJSON-part, they contain only
	a small number of detectable dependencies, between two and eleven. Therefore, 
	we refrain from computing the metrics precision and recall, as they are easily distorted when working with small numbers.
	However, the manual inspection of the derived dependencies, and the comparison against the target $\textit{Det}_\textit{Geo}$, shows that our approach is indeed successful for the input datasets chosen.
	
	To counter the threat of generalizability, further experiments over different real-world datasets (e.g., the datasets listed in Example~\ref{ex:other_datasets}), as well as synthetic data, and data not encoding any tagged unions, are required.
	
	We do not focus on the performance of our unoptimized prototype implementation, as the algorithm is only executed once for each dataset. 
	Of course, scalability to larger inputs is an issue to be addressed in future work, as CFD discovery in general is an expensive problem: 
	Already in addressing the discovery of functional dependencies in their full generality, algorithms have exponential runtime complexity~\cite{DBLP:journals/tkde/LiuLLC12}. 
	Several algorithms for the specific problem of CFD discovery have been proposed~\cite{4812508}. Their evaluations show that runtime performance depends heavily on the input, with some algorithms scaling better with size of the dataset (i.e., the number of tuples) thus being suitable for large datasets, while others perform better with higher arity. For constant CFDs, Li et al.~\cite{DBLP:journals/cj/LiLTY13} achieved promising runtime improvements by applying custom rules for
	pruning the search space.
	
	With strong restrictions to the problem space, such as our restriction to unary constant CFDs in our case, we can expect reasonable runtime performance on real-world inputs (which are often reasonably well-behaved).
	
	In our experiments, we observed a considerable impact of our heuristics and threshold on runtime, reducing the processing time for all datasets by up to two orders of magnitude. This is promising for further optimizations.

	\section{Conclusion and Outlook}
	\label{sec:future_work}
	\label{sec:conclusion}
	
	In this article, we proposed a method to infer conditional functional dependencies from JSON data.
	We use the identified dependencies 
	as the basis for declaring tagged unions in JSON schema extraction.
	
	This allows us to capture value-based constraints.
	In fact, in~\cite{DBLP:journals/is/GallinucciGR18} Gallinucci et al.\ report on expert interviews with users, regarding the users' preferences in schemas over nested data.
	Their interviews reveal that value-based conditions have a greater influence on the differentiation of schema variants than structural constraints, and are therefore preferred.

	In future work, our approach can be extended for the extraction of further variants of  dependencies, for instance, traditional conditional functional dependencies that capture implications between atomic property values, or dependencies where the value of the tag property implies the existence of a specific sibling property, as discussed in greater detail in Section~\ref{sec:extensions}. This allows to recognize a larger family of tagged unions.

	Our prototype implementation is currently main-mem\-ory based, which limits the size of inputs that we can handle.
	Making our implementation scale to larger inputs is one of the immediate next steps. Here, we may build upon first results by Mior~\cite{DBLP:journals/corr/Mior2021}, who shows that discovering dependencies in a relational encoding of JSON data has inferior runtime performance when compared to discovering dependencies in a streaming fashion.
	Also, we plan to consider a MapReduce-based approach to schema extraction, as implemented by Baazizi et al.\ in~\cite{DBLP:journals/vldb/BaaziziCGS19}.

	Ultimately, our goal is to obtain a schema declaration that human consumers consider to be comprehensive, but that  may also be efficiently processed programmatically. In order to obtain more succinct schemas, we need to resolve redundancies between the schema extracted by a third-party tool and our encoding of tagged unions.
	This requires rewriting the composite schema based on an algebraic representation of JSON Schema operators, such as the schema algebra proposed by Attouche et al.~\cite{DBLP:journals/corr/abs-2202-12849}.

	Further, identifying metrics that capture the quality of the extracted schemas will allow to quantitatively compare schemas extracted by different approaches. Given suitable metrics, the configuration of heuristics could even be adjusted automatically. A possible direction is to explore the notions of precision and recall, and the proxy-metric of schema entropy, as introduced by Spoth et al.\ in~\cite{DBLP:conf/sigmod/SpothKLHL21}.

	A further task is to adopt a CFD inference algorithm that is robust despite poor data quality and that can infer constraints despite outliers in the data. Naturally, this requires a relaxation to \enquote{soft} CFDs, a task where we may also build upon existing work on relational~\cite{DBLP:conf/pkdd/RammelaereG18} and even JSON data~\cite{DBLP:journals/corr/Mior2021}.

	In summary, our long-term vision is to extract comprehensible and therefore human-consumable schema declarations from JSON data. We believe that the detection of schema design patterns that are popular among schema designers, such as tagged unions, is an important building block towards realizing this vision.
	
	\paragraph{Acknowledgments:}
	This work was funded by \emph{Deut\-sche For\-schungs\-gemein\-schaft} (DFG, German Research Foundation) grant \#385808805.
	We thank Thomas Kirz for expertly typesetting the systems architecture in \LaTeX.
	
	\bibliographystyle{ACM-Reference-Format}
	\bibliography{main}
	
\end{document}